\documentclass[prc,manuscript,aps,superscriptaddress,showpacs]{revtex4}
\usepackage{amssymb}
\usepackage{amsmath,bm}
\usepackage{graphicx}
\usepackage[normalem]{ulem}
\usepackage[dvips]{color}

\renewcommand\sout{\bgroup \color{red} \ULdepth=-.5ex \ULset}

\begin{document}

\title{Density slope of the nuclear symmetry energy from the neutron skin thickness
of heavy nuclei}
\author{Lie-Wen Chen}
\affiliation{Department of Physics, Shanghai Jiao Tong University, Shanghai 200240, China}
\affiliation{Center of Theoretical Nuclear Physics, National Laboratory of Heavy Ion
Accelerator, Lanzhou 730000, China}
\author{Che Ming Ko}
\affiliation{Cyclotron Institute and Department of Physics and Astronomy, Texas A\&M
University, College Station, Texas 77843-3366, USA}
\author{Bao-An Li}
\affiliation{Department of Physics and Astronomy, Texas A\&M University-Commerce,
Commerce, Texas 75429-3011, USA}
\author{Jun Xu}
\affiliation{Cyclotron Institute and Department of Physics and Astronomy, Texas A\&M
University, College Station, Texas 77843-3366, USA}
\date{\today}

\begin{abstract}
Expressing explicitly the parameters of the standard Skyrme interaction in
terms of the macroscopic properties of asymmetric nuclear matter, we show in
the Skyrme-Hartree-Fock approach that unambiguous correlations exist between
observables of finite nuclei and nuclear matter properties. We find that
existing data on neutron skin thickness $\Delta r_{np}$ of Sn isotopes give
an important constraint on the symmetry energy $E_{\text{\textrm{sym}}}({%
\rho _{0}})$ and its density slope $L$ at saturation density ${\rho
_{0}}$. Combining these constraints with those from recent analyses
of isospin diffusion and double neutron/proton ratio in heavy-ion
collisions at intermediate energies leads to a more stringent limit
on $L$ approximately independent of $E_{\text{\textrm{sym}}}({\rho
_{0}})$. The implication of these new constraints on the $\Delta
r_{np}$ of $^{208}$Pb as well as the core-crust transition density
and pressure in neutron stars is discussed.
\end{abstract}

\maketitle

\section{Introduction}

The nuclear symmetry energy $E_{\text{\textrm{sym}}}(\rho )$ that encodes
the energy related to the neutron-proton asymmetry in the equation of state
(EOS) of isospin asymmetric nuclear matter (ANM) plays a crucial role in
both nuclear physics and astrophysics \cite%
{LiBA98,Dan02,Lat04,Ste05,Bar05,LCK08}. It is also relevant to some
interesting issues regarding possible new physics beyond the standard model
\cite{Hor01b,Sil05,Kra07,Wen09}. Although significant progress has been made
in recent years in determining the density dependence of $E_{\text{\textrm{%
sym}}}(\rho )$ \cite{Bar05,LCK08}, large uncertainties still exist even
around the normal density $\rho _{0}$. For instance, while the value of $E_{%
\text{\textrm{sym}}}(\rho_{0} )$ is determined to be around $30\pm 4$ MeV
mostly from analyzing nuclear masses, the extracted density slope $L$ of $E_{%
\text{\textrm{sym}}}(\rho )$ at $\rho _{0}$ scatters in a very large range
from about $20$ to $115$ MeV depending on the observables and methods used
in the studies \cite{Tsa09,Car10,She10}. Since many observables in
terrestrial laboratory experiments intrinsically depend on both $E_{\text{%
\textrm{sym}}}(\rho_{0} )$ and $L$, the extraction of $L$ at an accuracy
required for understanding more precisely many important properties of
neutron stars \cite{Lat04,XuJ09} is still severely prohibited, although the
uncertainty of $E_{\text{\textrm{sym}}}(\rho_{0} )$ is relatively small. To
extract $L$ with higher accuracy is thus of crucial importance.

Theoretically, studies based on both mean-field theories \cite%
{Bro00,Hor01,Fur02,Yos04,Che05b,Tod05,Rei10} and droplet-type models \cite%
{Mye69,Oya03,Dan03,Cen09} have shown that the neutron skin thickness $\Delta
r_{np}=\langle r_{n}^{2}\rangle ^{1/2}-\langle r_{p}^{2}\rangle ^{1/2} $ of
heavy nuclei, given by the difference of their neutron and proton
root-mean-squared radii, provides a good probe of $E_{\text{\textrm{sym}}%
}(\rho )$. In particular, $\Delta r_{np}$ has been found to correlate
strongly with both $E_{\text{\textrm{sym}}}(\rho_0 )$ and $L$ in microscopic
mean-field calculations \cite{Bro00,Hor01,Fur02,Yos04,Che05b,Tod05,Rei10}
using different parameter sets for the nuclear effective interactions, which
all fit the binding energies and charge radii of finite nuclei but
correspond to different $E_{\text{\textrm{sym}}}(\rho)$ and give different $%
\Delta r_{np}$. It is, however, difficult to extract an accurate value for $%
L $ from comparing calculated $\Delta r_{np}$ of heavy nuclei with
experimental data as $\Delta r_{np}$ depends on a number of nuclear
interaction parameters in a highly correlated manner \cite{Fur02,Rei10} and
the calculations have been usually carried out by varying simultaneously the
interaction parameters. Similar difficulties also exist when one tries to
extract other physical quantities from observables of finite nuclei within
mean-field theories or density functional theories~\cite{Col04}. A
well-known example is the Skyrme-Hartree-Fock (SHF) approach using normally $%
9$ interaction parameters. Although experimental data on
nucleon-nucleon scatterings and properties of both finite nuclei and
infinite nuclear matter would in principle put strong constraints on
the combinations of these
parameters~\cite{Vau72,Ton84,Mar02,Yos04,Agr05,Rei06,Cao06}, there
is generally no constraint on most of the individual interaction
parameters. Instead of varying directly the $9$ interaction
parameters within the SHF, we propose here an alternative approach
based on a modified Skyrme-Like (MSL) model \cite{Che09} to express
them explicitly in terms of $9$ macroscopic observables that are
either experimentally well constrained or empirically well known.
This opens the possibility to explore transparently the correlations
between properties of finite nuclei and the macroscopic properties
of nuclear matter within the SHF approach. In the present work, we
use this method to study the correlation between $\Delta r_{np}$ and
various macroscopic observables of infinite nuclear matter by
varying individually the values of the latter within their known
ranges. We then demonstrate that existing $\Delta r_{np}$ data on Sn
isotopes can give important constraints on $L$ and
$E_{\text{\textrm{sym}}}(\rho_0 )$. Combining these constraints with
those from recent analyses of isospin diffusion and double
neutron/proton ratio in heavy-ion collisions at intermediate
energies~\cite{Tsa09}, we further show that a more stringent
limit on $L$ is obtained approximately independent of the value of $E_{\text{%
\textrm{sym}}}({\rho _{0}})$. Finally, we discuss the implication of these
new constraints on both the $\Delta r_{np}$ of $^{208}$Pb
and the core-crust transition density and pressure in neutron
stars.

\section{The theoretical model}

\label{Theory}

In the present work, we use the so-called standard Skyrme interaction (see,
e.g., Ref.~\cite{Cha97}), which has been shown to be very successful in
describing the structure of finite nuclei, especially global properties such
as binding energies and charge radii, although non-standard extension is
possible~\cite{Cha97}. In the standard SHF model, the total energy density
of a nucleus is written as~\cite{Cha97}
\begin{equation}
\mathcal{H}=\mathcal{K}+\mathcal{H}_{0}+\mathcal{H}_{3}+\mathcal{H}_{eff}+%
\mathcal{H}_{fin}+\mathcal{H}_{SO}+\mathcal{H}_{sg}+\mathcal{H}_{Coul}
\label{HSky}
\end{equation}%
where $\mathcal{K}=\frac{\hbar ^{2}}{2m}\tau $ is the kinetic-energy term
and $\mathcal{H}_{Coul}$ is the Coulomb term, and $\mathcal{H}_{0}$, $%
\mathcal{H}_{3}$, $\mathcal{H}_{eff}$, $\mathcal{H}_{fin}$, $\mathcal{H}%
_{SO} $, $\mathcal{H}_{sg}$ are given by
\begin{eqnarray}
\mathcal{H}_{0} &=&t_{0}[(2+x_{0})\rho ^{2}-(2x_{0}+1)(\rho _{p}^{2}+\rho
_{n}^{2})]/4  \label{skyrme1} \\
\mathcal{H}_{3} &=&t_{3}\rho ^{\sigma }[(2+x_{3})\rho ^{2}-(2x_{3}+1)(\rho
_{p}^{2}+\rho _{n}^{2})]/24 \\
\mathcal{H}_{eff} &=&[t_{2}(2x_{2}+1)-t_{1}(2x_{1}+1)](\tau _{n}\rho
_{n}+\tau _{p}\rho _{p})/8  \notag \\
&&+[t_{1}(2+x_{1})+t_{2}(2+x_{2})]\tau \rho /8 \\
\mathcal{H}_{fin} &=&[3t_{1}(2+x_{1})-t_{2}(2+x_{2})](\nabla \rho )^{2}/32
\notag \\
&&-[3t_{1}(2x_{1}+1)+t_{2}(2x_{2}+1)]  \notag \\
&&\times \left[ (\nabla \rho _{n})^{2}+(\nabla \rho _{p})^{2}\right] /32
\label{Hfin0} \\
\mathcal{H}_{SO} &=&W_{0}[\overrightarrow{J}\cdot \overrightarrow{\nabla }%
\rho +\overrightarrow{J}_{p}\cdot \overrightarrow{\nabla }\rho _{p}+%
\overrightarrow{J}_{n}\cdot \overrightarrow{\nabla }\rho _{n}]/2 \\
\mathcal{H}_{sg} &=&(t_{1}-t_{2})[J_{p}^{2}+J_{n}^{2}]/16  \notag \\
&&-(t_{1}x_{1}+t_{2}x_{2})J^{2}/16  \label{skyrme7}
\end{eqnarray}%
in terms of the $9$ Skyrme interaction parameters $\sigma $, $t_{0}-t_{3}$, $%
x_{0}-x_{3}$, and the spin-orbit coupling constant $W_{0}$. In the above
equations, $\rho _{i}$, $\tau _{i}$ and $\vec{J}_{i}$ are, respectively, the
local nucleon number, kinetic energy and spin densities, whereas $\rho $, $%
\tau $ and $\vec{J}$ are corresponding total densities.

In the MSL model, the EOS of symmetric nuclear matter (SNM) and the nuclear
symmetry energy $E_{\text{\textrm{sym}}}(\rho )$ can be expressed,
respectively, as \cite{Che09}
\begin{equation}
E_{0}(\rho )=E_{\mathrm{kin}}^{0}u^{2/3}+Cu^{5/3}+\alpha u/2+\beta u^{\gamma
}/(\gamma +1),  \label{E0}
\end{equation}%
\begin{equation}
E_{\text{\textrm{sym}}}(\rho )=E_{\text{\textrm{sym}}}^{\mathrm{kin}}({\rho
_{0}})u^{2/3}+Du^{5/3}+E_{\text{\textrm{sym}}}^{\mathrm{loc}}({\rho })
\label{Esym}
\end{equation}%
where $u={\rho /}\rho _{0}$ is the reduced density; $E_{\mathrm{kin}}^{0}$
and $E_{\mathrm{sym}}^{\mathrm{kin}}$ are, respectively, the kinetic energy
at $\rho _{0}$ and its contribution to $E_{\text{\textrm{sym}}}(\rho )$; and
$E_{\text{\textrm{sym}}}^{\mathrm{loc}}({\rho })$ is the local
density-dependent symmetry energy given by%
\begin{equation}
E_{\text{\textrm{sym}}}^{\mathrm{loc}}({\rho })=(1-y)E_{\text{\textrm{sym}}%
}^{\mathrm{loc}}({\rho _{0}})u+yE_{\text{\textrm{sym}}}^{\mathrm{loc}}({\rho
_{0}})u^{\gamma }
\end{equation}%
with the dimensionless parameter%
\begin{equation}
y=\frac{L-3E_{\text{\textrm{sym}}}({\rho _{0}})+E_{\text{\textrm{\ sym}}}^{%
\mathrm{kin}}({\rho _{0}})-2D}{3(\gamma -1)E_{\text{ \textrm{sym}}}^{\mathrm{%
loc}}({\rho _{0}})}.
\end{equation}%
The model also includes following density-gradient term in the interaction
part of the binding energies for finite nuclei
\begin{equation}
E_{\mathrm{grad}}=G_{S}(\nabla \rho )^{2}/(2{\rho )}-G_{V}\left[ \nabla
(\rho _{n}-\rho _{p})\right] ^{2}/(2{\rho )},  \label{Egrad}
\end{equation}%
where $G_{S}$ and $G_{V}$ are the gradient and symmetry-gradient
coefficients~\cite{Ton84}.

By comparing expressions (\ref{E0}), (\ref{Esym}), and (\ref{Egrad}) in the
MSL model with corresponding ones in SHF, the $9$ Skyrme interaction
parameters in Eqs.~(\ref{skyrme1})-(\ref{skyrme7}) can be related to the $9$
parameters $\alpha $, $\beta $, $\gamma $, $C$, $D$, $E_{\text{\textrm{sym}}%
}^{\mathrm{loc}}({\rho _{0}})$, $y$, $G_{S}$ and $G_{V}$ in the MSL model by
following analytic relations
\begin{eqnarray}
t{_{0}} &=&4\alpha /(3{\rho _{0}}) \\
x{_{0}} &=&3(y-1)E_{\text{\textrm{sym}}}^{\mathrm{loc}}({\rho _{0}})/\alpha
-1/2 \\
t{_{3}} &=&16\beta /\left[ {\rho _{0}}^{\gamma }(\gamma +1)\right] \\
x{_{3}} &=&-3y(\gamma +1)E_{\text{\textrm{sym}}}^{\mathrm{loc}}({\rho _{0}}%
)/(2\beta )-1/2 \\
t_{1} &=&20C/\left[ 9{\rho _{0}(}k_{\mathrm{F}}^{0})^{2}\right] +8G_{S}/3 \\
t_{2} &=&\frac{4(25C-18D)}{9{\rho _{0}(}k_{\mathrm{F}}^{0})^{2}}-\frac{%
8(G_{S}+2G_{V})}{3} \\
x_{1} &=&\left[ 12G_{V}-4G_{S}-\frac{6D}{{\rho _{0}(}k_{\mathrm{F}}^{0})^{2}}%
\right] /(3t_{1}) \\
x_{2} &=&\left[ 20G_{V}+4G_{S}-\frac{5(16C-18D)}{3{\rho _{0}(}k_{\mathrm{F}%
}^{0})^{2}}\right] /(3t_{2}) \\
\text{\ }\sigma &=&\gamma -1
\end{eqnarray}%
with $k_{\mathrm{F}}^{0}=\left( 1.5\pi ^{2}{\rho _{0}}\right) ^{1/3}$. Since
the $7$ parameters $\alpha $, $\beta $, $\gamma $, $C$, $D$, $E_{\text{%
\textrm{sym}}}^{\mathrm{loc}}({\rho _{0}})$ and $y$ in the MSL model can be
expressed analytically in terms of the $7$ macroscopic quantities $\rho _{0}$%
, $E_{0}(\rho _{0})$, the incompressibility $K_{0}$, the isoscalar effective
mass $m_{s,0}^{\ast }$, the isovector effective mass $m_{v,0}^{\ast }$, $E_{%
\text{\textrm{sym}}}({\rho _{0}})$, and $L$ \cite{Che09}, the $9$ Skyrme
interaction parameters $\sigma $, $t_{0}-t_{3}$, $x_{0}-x_{3}$ can also be
expressed analytically in terms of the $9$ macroscopic quantities $\rho _{0}$%
, $E_{0}(\rho _{0})$, $K_{0}$, $m_{s,0}^{\ast }$, $m_{v,0}^{\ast }$, $E_{%
\text{\textrm{sym}}}({\rho _{0}})$, $L$, $G_{S}$, and $G_{V}$ via above
relations.
\begin{figure}[tbp]
\includegraphics[scale=1.2]{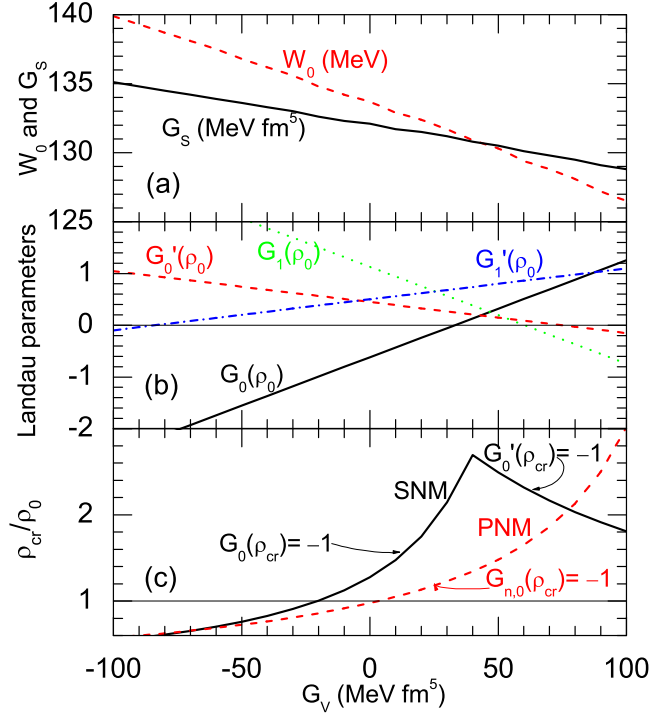}
\caption{{\protect\small (Color online) The symmetry-gradient coefficient }$%
G_{V}${\protect\small \ dependence of }$G_{S}${\protect\small \ and }$W_{0}$%
{\protect\small \ (a), Landau parameters }$G_{0}(\protect\rho _{0})$%
{\protect\small , }$G_{0}^{\prime }(\protect\rho _{0})${\protect\small , }$%
G_{1}(\protect\rho _{0})${\protect\small , }$G_{0}^{\prime }(\protect\rho %
_{0})${\protect\small \ (b), and the reduced critical density }$\protect\rho %
_{cr}/\protect\rho _{0}${\protect\small \ for symmetric nuclear matter and
pure neutron matter (c).}}
\label{LandauGV}
\end{figure}

As a reference for the correlation analyses below, we use following default
values for the macroscopic quantities. For the $7$ bulk properties of ANM,
we take $\rho _{0}=0.16$ fm$^{-3}$, $E_{0}(\rho _{0})=-16$ MeV, $K_{0}=230$
MeV, $m_{s,0}^{\ast }=0.8m$, $m_{v,0}^{\ast }=0.7m$, $E_{\text{\textrm{sym}}%
}({\rho _{0}})=30$ MeV, and $L=60$ MeV. Empirically, values of the gradient
coefficient $G_{S}$ and the symmetry-gradient coefficient $G_{V}$ are poorly
known, although they can be constrained by the
nuclear surface energy coefficient $a_{S}$, the Landau parameters for the
spin and spin-isospin channels in symmetric nuclear matter at the saturation
density, i.e., $G_{0}(\rho _{0})$, $G_{0}^{\prime }(\rho _{0})$, $G_{1}(\rho
_{0})$ and $G_{0}^{\prime }(\rho _{0})$, and the stability condition of
nuclear matter~\cite{Mar02,Agr05}. For the existing standard Skyrme
parameter sets, we have roughly $G_{S}=110\sim 150$ MeV$\cdot $fm$^{5}$ and $%
G_{V}=-40\sim 40$ MeV$\cdot $fm$^{5}$. Here we use the empirical value of
the surface energy coefficient $a_{S}=18$ MeV~\cite{Mye69} to determine $%
G_{S}$. We note that the surface energy coefficient $a_{S}$ also depends on
the spin-orbit coupling constant $W_{0}$~\cite{Tre86,Mar02,Agr05} that can
be determined by the neutron $p_{1/2}-p_{3/2}$ splitting in $^{16}$O. Since
the latter depends on $G_{S}$ and $G_{V}$, the three quantities $G_{S}$, $%
G_{V}$, and $W_{0}$ need to be determined simultaneously.

Keeping the $7$ bulk properties of ANM unchanged, we plot in Fig. \ref%
{LandauGV} (a) $G_{S}$ and $W_{0}$ as functions of $G_{V}$ with the values
of $G_{S}$, $W_{0}$, and $G_{V}$ simultaneously giving the surface energy
coefficient $a_{S}=18$ MeV and fitting the neutron $p_{1/2}-p_{3/2}$
splitting in $^{16}$O. In Figs. \ref{LandauGV} (b) and (c), we further show
the $G_{V}$ dependence of the Landau parameters $G_{0}(\rho _{0})$, $%
G_{0}^{\prime }(\rho _{0})$, $G_{1}(\rho _{0})$, $G_{0}^{\prime }(\rho _{0})$%
, and the critical density $\rho _{cr}$ above which at least one Landau
parameter violates the stability condition for symmetric nuclear matter and
pure neutron matter. Requiring the $\rho _{cr} $ of symmetric nuclear matter
to be larger than $\rho _{0}$ leads to $G_{V}\gtrsim -20$ MeV$\cdot $fm$^{5}$
and putting $\rho_{cr}>\rho _{0}$ for pure neutron matter further leads to $%
G_{V}\gtrsim 5$ MeV$\cdot $fm$^{5}$. Empirically, the Landau parameter $%
G_{0}^{\prime }$ has been extensively investigated, and its value can vary
from about zero to $1.6$ depending on the models and methods~\cite%
{Ost92,Bal98,Ben02,Zuo03,She03,Agr05,Wak05,Bor06}. In the present work with
the standard SHF approach, a positive $G_{0}^{\prime }$ leads to $G_{V}$ $%
\lesssim 70$ MeV$\cdot $fm$^{5}$ as shown in Fig. \ref{LandauGV} (b).
Furthermore, the Landau parameter $G_{0}^{\prime }$ can be extracted from
the spin-isospin response in finite nuclei, and its value has been found to
be $0.45\pm 0.06$ in the standard SHF approach~\cite{Fri86,Fra07}.
Therefore, we choose here $G_{V}=5$ MeV$\cdot $fm$^{5}$ which leads to $%
G_{0}^{\prime }=0.42$, $G_{S}=132$ MeV$\cdot $fm$^{5}$, and $W_{0}=133.3$ MeV%
$\cdot $fm$^{5}$. It is interesting to see that the value $G_{S}=132$ MeV$%
\cdot $fm$^{5}$ is quite consistent with that used extensively in the
literature~\cite{Oya03,Oya98,XuJ09}. This new Skyrme parameter set obtained
with above empirical values for the macroscopic quantities is referred as
MSL0. Summarized in Table \ref{MSL0} are values of corresponding Skyrme
parameters and some macroscopic quantities.

\begin{table}[tbp]
\caption{{\protect\small {Skyrme parameters in MSL0 (left side) and some
corresponding nuclear properties (right side).}}}
\label{MSL0}%
\begin{tabular}{lr||lr}
\hline\hline
Quantity & MSL0 & Quantity & MSL0 \\ \hline
$t_{0}$ (MeV$\cdot $fm$^{3}$) & $-2118.06$ & $\rho _{0}$ (fm$^{-3}$) & $0.16$
\\
$t_{1}$ (MeV$\cdot $fm$^{5}$) & $395.196$ & $E_{0}$ (MeV) & $-16.0$ \\
$t_{2}$ (MeV$\cdot $fm$^{5}$) & $-63.9531$ & $K_{0}$ (MeV) & $230.0$ \\
$t_{3}$ (MeV$\cdot $fm$^{3+3\sigma }$) & $12857.7$ & $m_{s,0}^{\ast }/m$ & $%
0.80$ \\
$x_{0}$ & $\ -0.0709496$ & $m_{v,0}^{\ast }/m$ & $0.70$ \\
$x_{1}$ & $-0.332282$ & $E_{\text{sym}}(\rho _{0})$ (MeV) & $30.0$ \\
$x_{2}$ & $1.35830$ & $L$ (MeV) & $60.0$ \\
$x_{3}$ & $-0.228181$ & $G_{S}$ (MeV$\cdot $fm$^{5}$) & $132.0$ \\
$\sigma $ & $0.235879$ & $G_{V}$ (MeV$\cdot $fm$^{5}$) & $5.0$ \\
$W_{0}$ (MeV$\cdot $fm$^{5}$) & $133.3$ & $G_{0}^{\prime }(\rho _{0})$ & $%
0.42$ \\ \hline\hline
\end{tabular}%
\end{table}

\begin{figure}[tbp]
\includegraphics[scale=1.2]{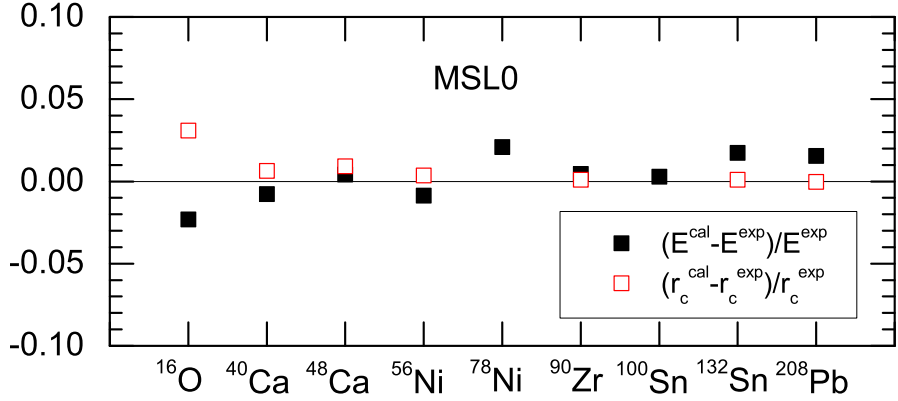}
\caption{{\protect\small (Color online) Relative deviation of the binding
energies and charge rms radii of }$^{16}${\protect\small O, }$^{40}$%
{\protect\small Ca, }$^{48}${\protect\small Ca, }$^{56}${\protect\small Ni, }%
$^{78}${\protect\small Ni, }$^{90}${\protect\small Zr, }$^{100}$%
{\protect\small Sn, }$^{132}${\protect\small Sn, and }$^{208}$%
{\protect\small Pb from SHF with MSL0.}}
\label{EbRc}
\end{figure}

To test the new Skyrme parameter set MSL0, we calculate the binding energies
and charge rms radii for a number of closed-shell or semi-closed-shell
nuclei: $^{16}$O, $^{40}$Ca, $^{48}$Ca, $^{56}$Ni, $^{78}$Ni, $^{90}$Zr, $%
^{100}$Sn, $^{132}$Sn, and $^{208}$Pb. Figure \ref{EbRc} shows the
relative deviation of the charge rms radii and binding energies of
these nuclei from those measured in
experiments~\cite{Aud03,Ang04,LeB05}. It is seen that the MSL0 can
describe the experimental data very well except for the light
nucleus $^{16}$O for which the deviation reaches to about $2-3\%$.
This is a remarkable result as MSL0 is not obtained from fitting
measured binding energies and charge rms radii of finite nuclei as
in usual Skyrme parametrization. It should be pointed out that our
main motivation for introducing the MSL0 is not to construct another
Skyrme parameter set to describe data, but to use as a reference for
the correlation analyses in the following. As we will show, varying
$G_{S}$, $G_{V}$ and $W_{0}$ will not affect the conclusion in the
present work.

\section{Results}

\label{Result}

\begin{figure}[tbp]
\includegraphics[scale=1.2]{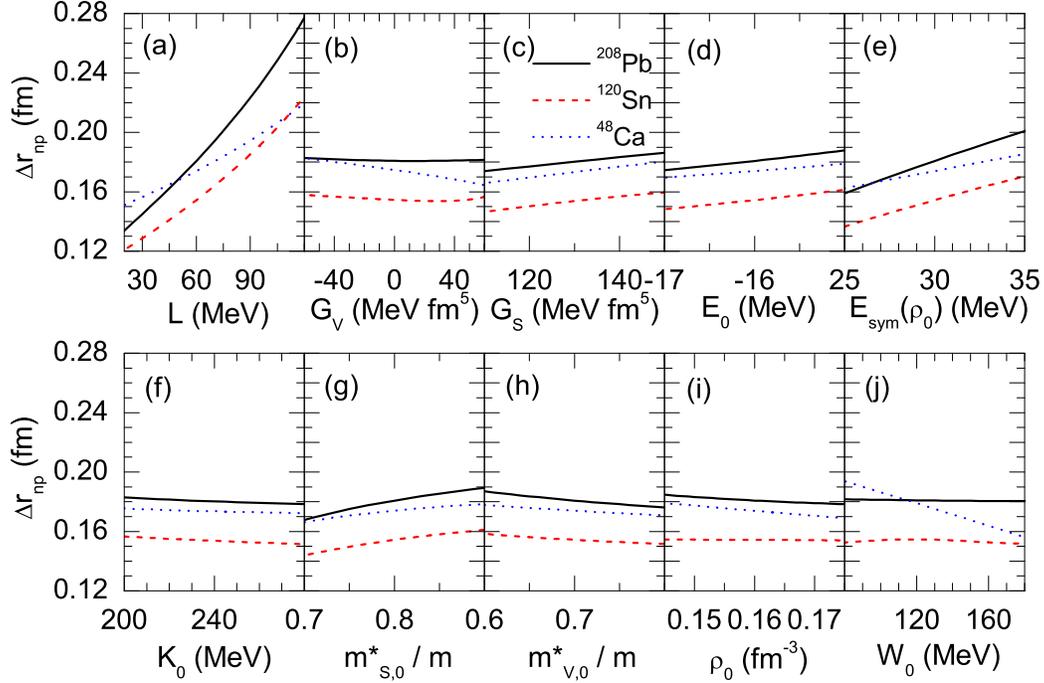}
\caption{{\protect\small (Color online) The neutron skin thickness }$\Delta
r_{np}${\protect\small \ of }$^{208}${\protect\small Pb, }$^{120}$%
{\protect\small Sn and }$^{48}${\protect\small Ca from SHF with MSL0 by
varying individually }$L${\protect\small \ (a), }$G_{V}${\protect\small \ (b), }$G_{S}$%
{\protect\small \ (c), }$E_{0}(\protect\rho _{0})${\protect\small \ (d), }$E_{\text{%
\textrm{sym}}}(\protect\rho _{0})${\protect\small \ (e), }$K_{0}${\protect\small \ (f)%
, }$m_{s,0}^{\ast }${\protect\small \ (g), }$m_{v,0}^{\ast }${\protect\small \ (h), }$%
\protect\rho _{0}${\protect\small \ (i), and }$W_{0}${\protect\small
\ (j).}} \label{RnpPbSnCa}
\end{figure}

To reveal clearly the dependence of $\Delta r_{np}$ on each macroscopic
quantity, we vary one quantity at a time while keeping all others at their
default values in MSL0. Shown in Fig. \ref{RnpPbSnCa} are the values of $%
\Delta r_{np}$ for $^{208}$Pb, $^{120}$Sn and $^{48}$Ca. Within the
uncertain ranges considered here, the $\Delta r_{np}$ of $^{208}$Pb and $%
^{120}$Sn exhibits a very strong correlation with $L$. However, it depends
only moderately on $E_{\text{\textrm{sym}}}({\rho _{0}})$ and weakly on $%
m_{s,0}^{\ast }$. On the other hand, the $\Delta r_{np}$ of $^{48}$Ca
displays a much weaker dependence on both $L$ and $E_{\text{\textrm{sym}}}({%
\rho _{0}})$. Instead, it depends moderately on $G_{V}$ and $W_{0}$. This
explains the weaker $\Delta r_{np}$-$E_{\text{\textrm{sym}}}({\rho })$
correlation observed for $^{48}$Ca in previous SHF calculations using
different interaction parameters \cite{Che05b}. These results demonstrate
that the $\Delta r_{np}$ of heavy nuclei can provide reliable information on
the symmetry energy around the normal density.

\begin{figure}[tbp]
\includegraphics[scale=1.2]{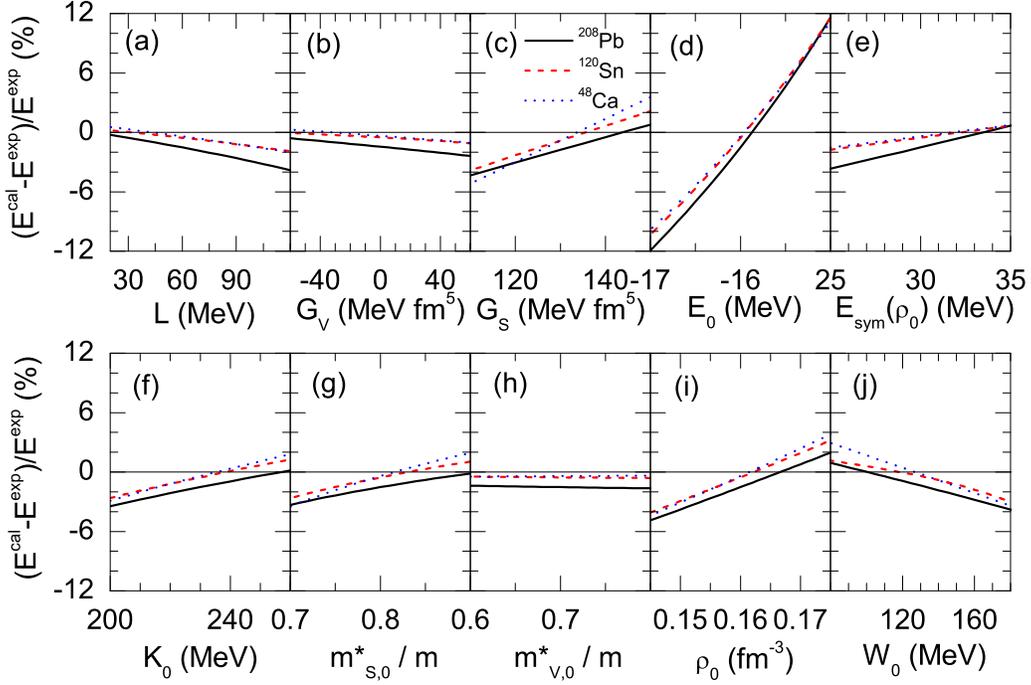}
\caption{{\protect\small (Color online) Relative deviation of the binding
energies of }$^{208}${\protect\small Pb, }$^{120}${\protect\small Sn and }$%
^{48}${\protect\small Ca from SHF with MSL0 by varying individually }$L$%
{\protect\small \ (a), }$G_{V}${\protect\small \ (b), }$G_{S}${\protect\small \ (c), }$%
E_{0}(\protect\rho _{0})${\protect\small \ (d), }$E_{\text{\textrm{sym}}}(\protect%
\rho _{0})${\protect\small \ (e), }$K_{0}${\protect\small \ (f), }$m_{s,0}^{\ast }$%
{\protect\small \ (g), }$m_{v,0}^{\ast }${\protect\small \ (h), }$\protect\rho _{0}$%
{\protect\small \ (i), and }$W_{0}${\protect\small \ (j).}}
\label{EbPbSnCa}
\end{figure}

\begin{figure}[tbp]
\includegraphics[scale=1.2]{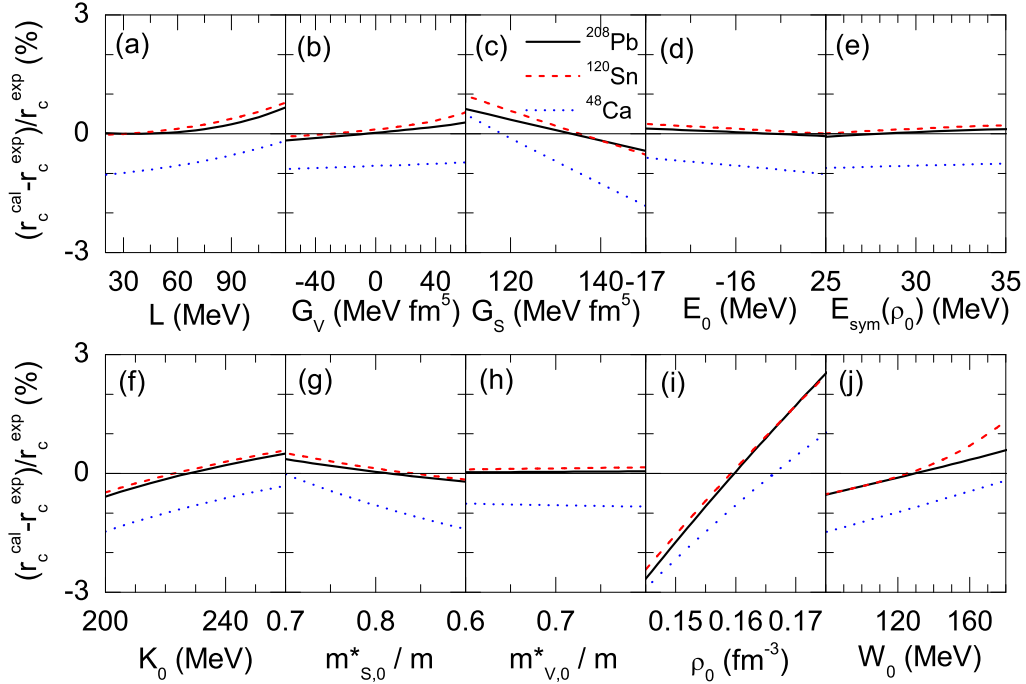}
\caption{{\protect\small (Color online) Same as Fig.~\protect\ref{EbPbSnCa}
but for the charge rms radii.}}
\label{RcPbSnCa}
\end{figure}

As we vary one of the macroscopic quantities in Fig.
\ref{RnpPbSnCa}, it is of interest to see how this affects the
binding energy and charge rms radius. This is shown in Fig.
\ref{EbPbSnCa} and Fig. \ref{RcPbSnCa}, respectively, for the
changes in the relative deviation of the binding energies and charge
rms radii of $^{208}$Pb, $^{120}$Sn and $^{48}$Ca from the data.
Within the uncertain ranges for the macroscopic quantities
considered here, it is seen that the relative deviation of the
binding energy is basically less than $3\%$ except the case for the
macroscopic quantity $E_0$. For the charge rms radius, the relative
deviation is even much smaller. Especially, for the heavy $^{208}$Pb
and $^{120}$Sn, the relative deviation of the charge rms radius is
basically less than $0.5\%$ except the case for the macroscopic
quantity $\rho_0$. These features imply that the binding energies
and charge rms radii of finite nuclei can be
reasonably reproduced when we perform the correlation analysis shown in Fig. %
\ref{RnpPbSnCa} by varying individually the macroscopic quantity. This is
particularly the case when we only change the macroscopic quantities $L $
and $E_{\text{\textrm{sym}}}({\rho _{0}})$ from the MSL0 in the correlation
analyses.

\begin{figure}[tbp]
\includegraphics[scale=1.5]{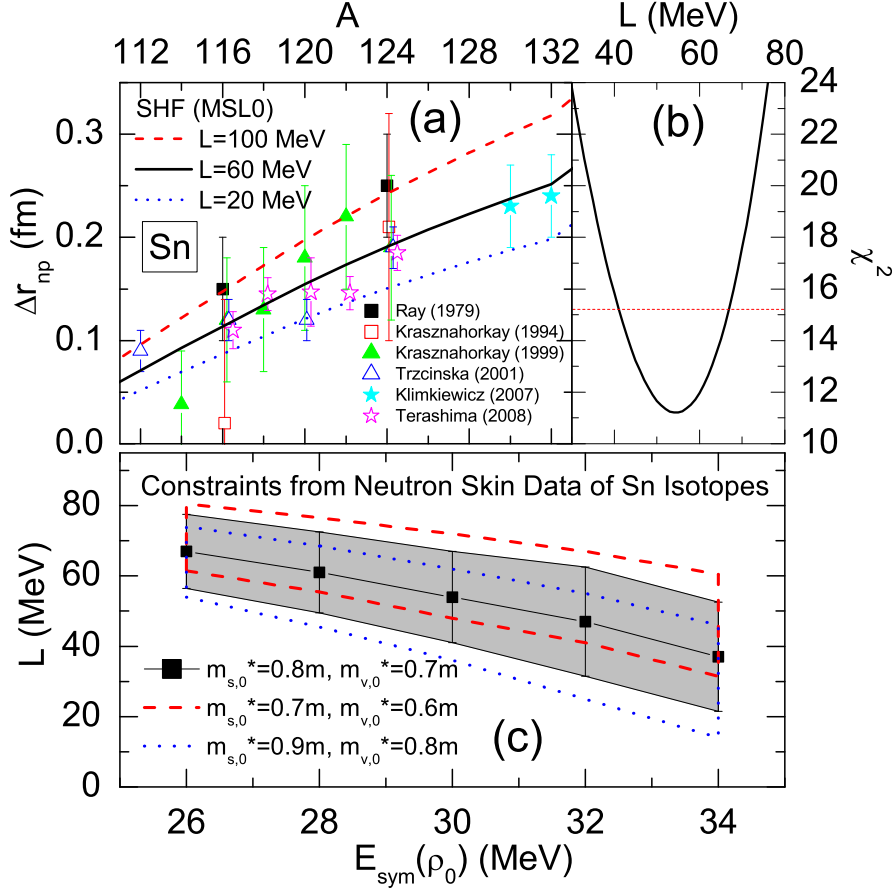}
\caption{{\protect\small (Color online) (a): The }$\Delta r_{np}
${\protect\small \ data for Sn isotopes from different experimental
methods and results from SHF calculation using MSL0 with }$L=20$%
{\protect\small , }$60${\protect\small \ and }$100${\protect\small \
MeV. (b): }$\protect\chi ^{2}${\protect\small \ as a function of }$L
${\protect\small . (c): Constraints on }$L${\protect\small \ and }$%
E_{\text{\textrm{sym}}}(\protect\rho _{0})${\protect\small \ from the }$%
\protect\chi ^{2}${\protect\small \ analysis of the }$\Delta r_{np}$%
{\protect\small \ data on Sn isotopes (Grey band as well as dashed and
dotted lines). }}
\label{RnpSnL}
\end{figure}

Experimentally, much effort has been devoted to determining the values of $%
\Delta r_{np}$ for finite nuclei using various methods. In particular, the $%
\Delta r_{np}$ of heavy Sn isotopes has been systematically measured \cite%
{Ray79,Kra94,Kra99,Trz01,Kli07,Ter08}. As an illustration, we first show in
Fig. \ref{RnpSnL} (a) the comparison of available Sn $%
\Delta r_{np}$ data with our calculated results using different values of $%
20 $, $60$ and $100$ MeV for $L$ and the default values for all
other quantities in MSL0. It is seen that the value $L=60$ MeV best
describes the data. To be more precise, the $\chi ^{2}$ evaluated
from the difference between the theoretical and experimental $\Delta
r_{np}$ values is shown as a function of $L$ in Fig. \ref{RnpSnL}
(b). The most reliable value of $L$ is found to be $L=54\pm 13$ MeV
within a $2\sigma $ uncertainty.

Since the value of $\Delta r_{np}$ depends on both $L$ and $E_{\text{\textrm{%
sym}}}({\rho _{0}})$, a two-dimensional $\chi ^{2}$ analysis as
shown by the grey band in Fig. \ref{RnpSnL} (c) is necessary. It is
seen that increasing the value of $E_{\text{\textrm{sym}}}({\rho
_{0}})$ systematically leads to smaller values of $L$. More
quantitatively, the value of $L$ varies from $67\pm 10.5$ to $37\pm
15.5$ MeV when the value of $E_{\text{\textrm{sym}}}({\ \rho _{0}})$
changes from $26$ to $34$ MeV. Furthermore, we have estimated the
effects of nucleon effective mass by using $m_{s,0}^{\ast }=0.7m$
and $m_{v,0}^{\ast }=0.6m$ as well as $m_{s,0}^{\ast }=0.9m$ and
$m_{v,0}^{\ast }=0.8m$, in accord with the empirical constraint
$m_{s,0}^{\ast }>m_{v,0}^{\ast }$ \cite{LCK08,Les06}, and the
resulting constraints are shown by the dashed and dotted lines. As
expected from the results shown in Fig.\ \ref{RnpPbSnCa}, effects of
nucleon effective mass are small with the value of $L$ shifting by
only a few MeV for a given $E_{\text{\textrm{sym}}}( {\rho _{0}})$.
We have also checked that effects of varying other macroscopic
quantities are even smaller.

\begin{figure}[tbp]
\includegraphics[scale=1.2]{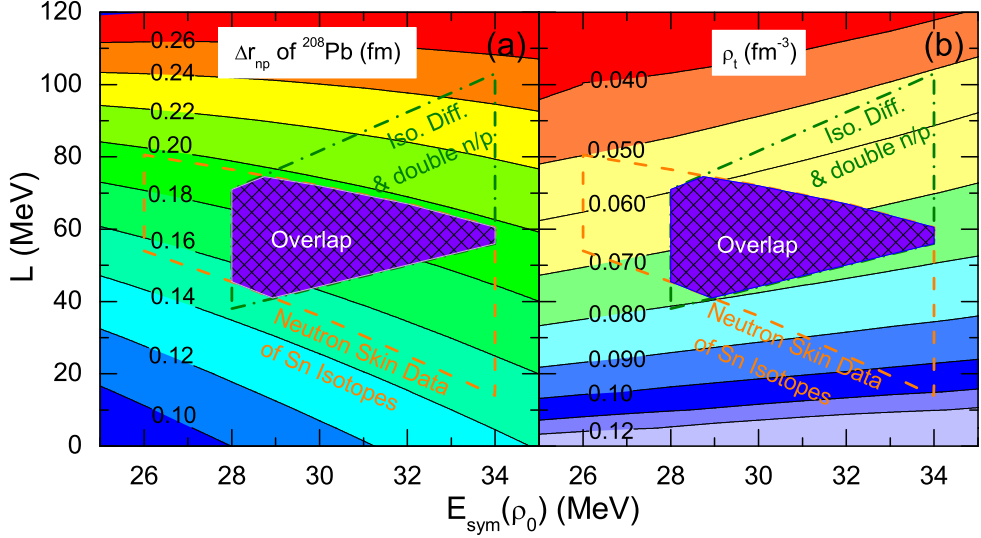}
\caption{{\protect\small (Color online) Contour curves in the }$E_{\text{%
\textrm{sym}}}(\protect\rho _{0})${\protect\small -}$L${\protect\small \
plane for the }$\Delta r_{np}${\protect\small \ of }$^{208}${\protect\small %
Pb (a) from SHF calculation with MSL0 and the core-crust transition
density $\protect\rho _{t}$ (b). The shaded region represents the
overlap of constraints obtained in the present work (dashed lines)
and that from Ref.\ \protect\cite{Tsa09} (dash-dotted lines).}}
\label{rhoTPb}
\end{figure}

The above constraints on the $L$-$E_{\text{\textrm{sym}}}({\rho
_{0}})$ correlation can be combined with those from recent analyses
of isospin diffusion and double $n/p$ ratio in heavy-ion collisions
at intermediate energies \cite{Tsa09} to determine simultaneously
the values of both $L$ and $E_{\text{\textrm{sym}}}({\rho _{0}})$.
Shown in Fig. \ref{rhoTPb} (a) (and (b)) are the two constraints in
the $E_{\text{\textrm{sym}}}({\rho _{0}})$-$L$ plane. Interestingly,
these two constraints display opposite
$L$-$E_{\text{\textrm{sym}}}({\rho _{0}})$ correlations. This allows
us to extract a value of $L=58\pm 18$ MeV approximately independent
of the value of $E_{\text{\textrm{sym}}}({\rho _{0}})$. This value
of $L$ is essentially overlapped with other constraints extracted
from different experimental data in the
literature~\cite{Tsa09,Car10,She10} but with much higher precision
although the constraint on $E_{\text{\textrm{sym}}}({\rho _{0}})$ is
not improved. It also agrees well with the value of $L=66.5$ MeV
obtained from a recent systematic analysis of the density dependence
of nuclear symmetry energy within the microscopic
Brueckner-Hartree-Fock approach using the realistic Argonne V18
nucleon-nucleon potential plus a phenomenological three-body force
of Urbana type~\cite{Vid09}. Furthermore, it is in remarkably good
agreement with the value of $L=52.7$ MeV extracted most recently
from global nucleon optical potentials constrained by world data on
nucleon-nucleus and (p,n) charge-exchange reactions~\cite{XuC10}.

Also shown in Fig. \ref{rhoTPb} (a) are contours of the $\Delta
r_{np}$ of $^{208}$Pb. Based on the constraints on $L$ and
$E_{\text{\textrm{sym}}}({\rho _{0}})$ shown by the shaded region in
Fig. \ref{rhoTPb}, it is seen that the $\Delta r_{np}$ of $^{208}$Pb
is tightly limited to a narrow region of $0.175\pm 0.02$ fm, which
is quite consistent with other constraints from various experiments
\cite{LCK08} but with much smaller uncertainty. The Lead Radius
Experiment (PREX) \cite{PREX} being preformed at Jefferson Lab aims
to determine model-independently the $\langle r_{n}^{2}\rangle
^{1/2}$ of $^{208}$Pb to $1\%$ accuracy, and this is expected to
further improve the determination of $E_{\text{\textrm{sym}}}(\rho
)$ at subnormal densities.

\begin{figure}[tbp]
\includegraphics[scale=1.2]{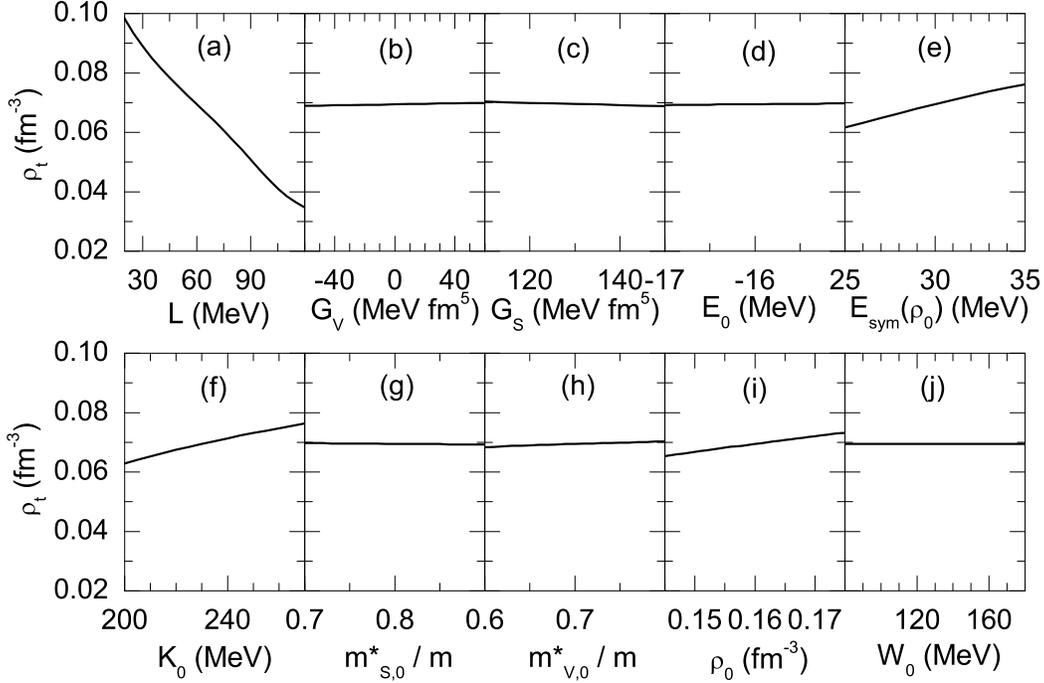}
\caption{{\protect\small Same as Fig.~\protect\ref{RnpPbSnCa}
but for the core-crust transition density }$\protect\rho _{t}$%
{\protect\small \ in neutron stars.}}
\label{XrhoT}
\end{figure}

\begin{figure}[tbp]
\includegraphics[scale=1.2]{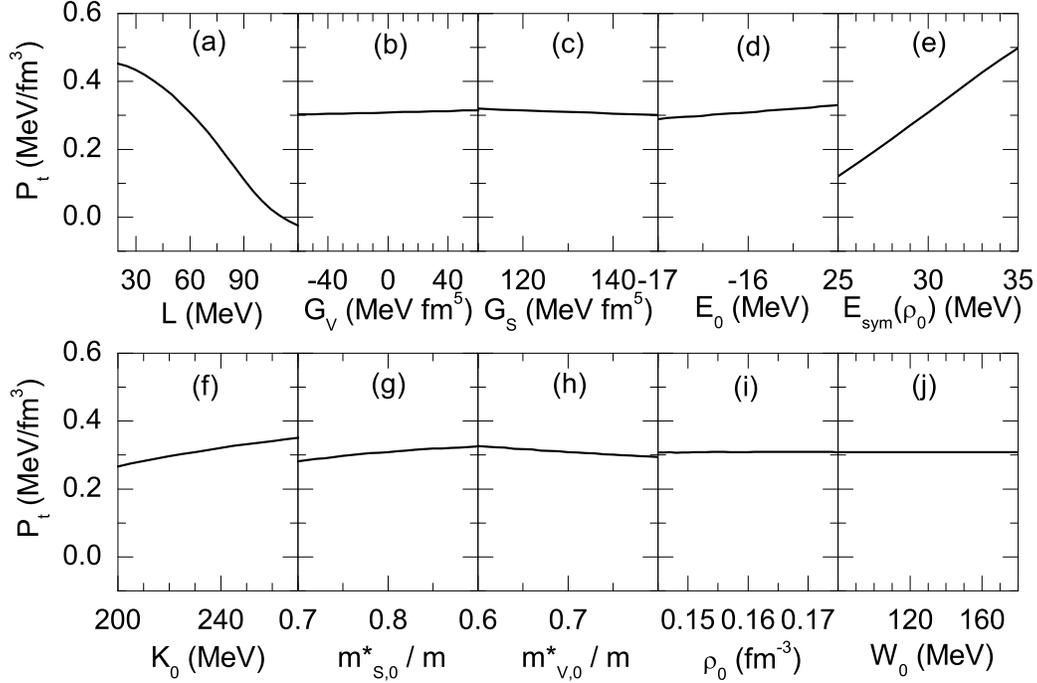}
\caption{{\protect\small Same as Fig.~\protect\ref{RnpPbSnCa} but
for the the core-crust transition pressure }$P_{t}${\protect\small \
in neutron stars.}} \label{XpT}
\end{figure}

To see the implications of our results in astrophysics, we have
carried out a similar correlation analysis for the transition
density $\rho _{t}$ and corresponding pressure $P_{t}$ at the inner
edge of neutron star crusts, which play crucial roles in neutron
star properties~\cite{Lat04,XuJ09}, using their values evaluated in
a dynamical approach \cite{XuJ09}, and the results are shown in Fig.
\ref{XrhoT} and Fig. \ref{XpT}, respectively. It is seen that the
$\rho _{t}$ ($P_{t}$) displays a particularly strong correlation
with $L$ ($L$ and $E_{\text{\textrm{sym}}}({\rho _{0}})$), a weak
dependence on $E_{\text{\textrm{sym}}}({\rho _{0}})$ and $K_{0}$
($K_{0} $), but almost no sensitivity to other macroscopic
parameters. These features are consistent with the results in
Ref.~\cite{XuJ09} where $\rho _{t}$ has been shown to display a
stronger correlation with $L$ than $P_{t}$ in SHF calculations using
different interaction parameters. The contours of the core-crust
transition density $\rho _{t}$ in neutron stars in the
$E_{\text{\textrm{sym}}}({\rho _{0}})$-$L$ plane is shown in Fig.
\ref{rhoTPb} (b). It shows that the value of $\rho _{t}$ is limited
to $0.069\pm 0.011$ fm$^{-3}$ by the constraints on $L$ and
$E_{\text{\textrm{sym}}}({\rho _{0}})$ obtained in the present work.
Including further the uncertainty in the value of $K_{0}$, we obtain
a value of $\rho _{t}=0.069\pm 0.018$ fm$^{-3}$. A similar analysis
leads to $P_{t}=0.33\pm 0.21$ MeV/fm$^{3}$. These results agree well
with the empirical values \cite{Lat04} but are slightly larger than
previous results in Ref. \cite{XuJ09} using
$E_{\text{\textrm{sym}}}({\rho _{0}})=30.5$ MeV and $L=86\pm 25$ MeV
extracted only from the isospin diffusion data in heavy-ion
collisions \cite{Che05a}.

\section{Summary}

\label{Summary}

We have proposed to analyze the correlation between observables of finite
nuclei and some macroscopic properties of asymmetric nuclear matter by
expressing explicitly the parameters of the nuclear effective interaction in
terms of the macroscopic properties of asymmetric nuclear matter. This would
allow us to extract information on some important physical quantities from
data on finite nuclei in a more transparent way.

Using such a correlation analysis within the standard SHF approach, we have
demonstrated that the neutron skin thickness of heavy nuclei can provide
reliable information on the symmetry energy, and the existing neutron skin
data on Sn isotopes can give important constraints on the symmetry energy
parameters $E_{\text{\textrm{sym}}}({\rho _{0}})$ and $L$. In particular,
combining the obtained $L$-$E_{\text{\textrm{sym}}}({\rho _{0}})$
constraints with that from recent analyses of isospin diffusion and double $%
n/p$ ratio in heavy-ion collisions has led to a quite accurate value of $%
L=58\pm 18$ MeV approximately independent of the value of $E_{\text{\textrm{%
sym}}}({\rho _{0}})$. The obtained $L$-$E_{\text{\textrm{sym}}}({\rho _{0}})$
constraints also put a stringent limit of $\Delta r_{np}=0.175\pm 0.02$ fm
for the neutron skin thickness of $^{208}$Pb.

Furthermore, we have explored how the core-crust transition density
$\rho _{t}$ and the corresponding pressure $P_{t}$ in neutron stars
correlate with the macroscopic properties of asymmetric nuclear
matter. Our results have indicated that the $\rho _{t}$ displays a
particularly strong correlation with $L$, a weak dependence on $E_{\text{%
\textrm{sym}}}({\rho _{0}})$ and $K_{0}$, but almost no sensitivity to other
macroscopic parameters. On the other hand, the $P_{t} $ exhibits a strong
correlation with both $L$ and $E_{\text{\textrm{sym}}}({\rho _{0}})$, and a
weak dependence on $K_{0} $. The $L$-$E_{\text{\textrm{sym}}}({\rho _{0}})$
constraints obtained in the present work leads to $\rho _{t}=0.069\pm 0.018$
fm$^{-3}$ and $P_{t}=0.33\pm 0.21$ MeV/fm$^{3}$.

Although we have mainly analyzed in the present work the correlation
between the neutron skin thickness of finite nuclei and the symmetry
energy within the standard SHF approach, our method can be
generalized to other correlation analyses or mean-field models. In
particular, it will be interesting to see how our results will
change if different energy density functions are used. These studies
are in progress.

\section*{ACKNOWLEDGMENTS}

This work was supported in part by the NNSF of China under Grant No.
10975097, Shanghai Rising-Star Program under Grant No. 06QA14024, the
National Basic Research Program of China (973 Program) under Contract No.
2007CB815004 and 2010CB833000, U.S. NSF under Grant No. PHY-0758115 and
PHY-0757839, the Welch Foundation under Grant No. A-1358, the Research
Corporation under Award No. 7123, the Texas Coordinating Board of Higher
Education Award No. 003565-0004-2007.

\end{document}